\newcommand{\version}{January 18, 2022}
\documentclass[letterpaper]{twocolartcl}

\renewcommand{\d}{\delta}
\newcommand{\e}{\epsilon}

\newcommand{\vth}{\vartheta}

\renewcommand{\l}{\lambda}

\newcommand{\rh}{\rho}
\newcommand{\s}{\sigma}

\newcommand{\ph}{\phi}

\newcommand{\w}{\omega}

\newcommand{\G}{\Gamma}

\newcommand{\W}{\Omega}

  \newcommand{\cO}{\mathcal{O}}

\newcommand{\pa}{\partial}

\newcommand{\nn}{\nonumber}
\newcommand{\eqnref}[1]{Eq. \eqref{#1}}

\newcommand{\bfw}{\mathbf{w}}

\newcommand{\ct}{c_{\textrm{T}}}
\newcommand{\cl}{c_{\textrm{L}}}
\newcommand{\cs}{c_{\textrm{S}}}

\newcommand{\bt}{\beta_{\textrm{T}}}

\newcommand{\vc}{v_{\textrm{c}}} 

\newcommand{\vq}{\vv{q}}
\newcommand{\vqi}{\vv{q}\,^{\!'}}
\newcommand{\vqii}{\vv{q}\,^{\!''}}

\titlehead{\vspace{-2em}\hfill \version\\
\phantom{ }\hfill LA-UR-19-20476}
       \title{\texorpdfstring{\vspace{-1.2em}}{}%
       Properties of dislocation drag from phonon wind at ambient conditions}

\author{Daniel N. Blaschke}

\date{\large Los Alamos National Laboratory, Los Alamos, NM, 87545, USA
\\\ttfamily{E-mail: dblaschke@lanl.gov}}

\newcommand{\keywords}{dislocations in crystals, drag coefficient, phonon wind}
\hypersetup{pdfkeywords = {\keywords}}

\graphicspath{{./}{./figures/}}

\begin{document}

\maketitle

\begin{abstract}
It is well known that under plastic deformation, dislocations are not only created but also move through the crystal,
and their mobility is impeded by their interaction with the crystal structure.
At high stress and temperature, this ``drag'' is dominated by phonon wind, i.e. phonons scattering off dislocations.
Employing the semi-isotropic approach discussed in detail in Ref.~\cite{Blaschke:2018anis}, we discuss here the approximate functional dependence of dislocation drag $B$ on dislocation velocity in various regimes between a few percent of transverse sound speed $\ct$ and $\ct$ (where $\ct$ is the effective average transverse sound speed of the polycrystal).
In doing so, we find an effective functional form for dislocation drag $B(v)$ for different slip systems and dislocation characters at fixed (room) temperature and low pressure.
\end{abstract}

\vspace{-0.2cm}
\toc
\vspace{-0.2cm}

\section{Introduction}
\label{sec:intro}

Many modern material strength models, such as for example those of Refs.~\cite{Krasnikov:2010,Barton:2011,Hansen:2013,Hunter:2015,Borodin:2015,Luscher:2016,Austin:2018}, are based on dislocation dynamics.
Yet, dislocation mobility, especially in the high temperature and high stress regime, is poorly understood theoretically.
Moving dislocations experience a drag due to their interaction with the crystal structure, and this drag coefficient $B$ determines the dislocation glide time between obstacles.
The lack of a well-established functional form for $B(v,T,\ldots)$ has led many researchers to assume $B$ to be a constant (or a constant over a simple ``relativistic'' factor) as a fist order approximation within their strength models.
Thus, better insight into the true functional form of $B$ could improve those models.

Different mechanisms dominate dislocation drag in different regimes. However, at temperatures comparable to or higher than the Debye temperature and at high stress (leading to dislocation velocities in the range $0.01\lesssim v/\cs<1$, where $\cs$ denotes the lowest shear wave speed corresponding to the direction of dislocation glide),
phonons scattering off dislocations (commonly referred to as ``phonon wind'') constitutes the dominating effect.
The lower end of this range is known as the ``viscous'' regime where $B(v)$ at given temperature and pressure is known to be roughly constant.
However, with increasing stress and thus increasing dislocation velocity, $B$ exhibits a non-linear velocity dependence.
This is seen in numerous molecular dynamics (MD) simulations, see e.g.~\cite{Olmsted:2005,Marian:2006,Wang:2008,Gilbert:2011,Daphalapurkar:2014} and references therein, but also within the recent theoretical framework of Refs.~\cite{Blaschke:BpaperRpt,Blaschke:2018anis}.

In Ref.~\cite{Blaschke:2018anis}, the theory developed by Alshits and collaborators~\cite{Alshits:1992} was taken to the next level by including not only the full velocity dependence of $B$, but by also including longitudinal phonons (in addition to the dominating contribution of transverse phonons) as well as an anisotropic dislocation field and single crystal elastic constants.
Hence, this model which was developed having polycrystals in mind, keeps the phonon spectrum isotropic (for simplicity), but dislocations are modeled according to the single crystal symmetry (bcc, fcc, hcp, etc.) in order to take into account their anisotropy to some extent.
This ``semi-isotropic'' approach constitutes an intermediate step in an ongoing long-term endeavor to include all anisotropic effects and the true phonon spectrum (which is beyond the scope of the current work).
Nonetheless, valuable insights were already gained, like the non-trivial dependence of the drag coefficient on the dislocation character angle $\vth$ (between line sense and Burgers vector).

For now, the model is also restricted to the subsonic regime where $v<\ct$.
The question whether dislocations in metals can reach supersonic speeds is still under debate, although
numerous MD simulations suggest it is possible~\cite{Rosakis:2001,Li:2002,Olmsted:2005,Jin:2008,Gilbert:2011,Ruestes:2015}; see also the recent discussion on interpreting those results in the context of line tension and dislocation shape~\cite{Blaschke:2017lten}.
For recent literature on supersonic dislocations, see e.g.~\cite{Markenscoff:2009,Pellegrini:2010,Pellegrini:2017} and references therein.

Here, our main goal is to highlight the effective functional dependence of $B$ on the dislocation velocity within the theory of~\cite{Blaschke:2018anis} (and its numerical implementation of Ref.~\cite{pydislocdyn}), and to explain how to derive simple analytic representations of $B(v)$ which are amenable to subsequent use in applications (such as material strength models).
We also present new results for metals and slip systems not presented in~\cite{Blaschke:2018anis} (i.e. prismatic and pyramidal slip for hcp metals).
Thus, the current work in a sense complements Ref.~\cite{Blaschke:2018anis} and the theory developed there.

\section{Phonon wind in the semi-isotropic approach}
\label{sec:phononwind}

The drag coefficient $B$ of a dislocation is defined as the proportionality coefficient of the force $F=Bv$ needed to maintain dislocation velocity $v$.
It is related to the dissipation $D$ per unit length due to phonon scattering via $D=Bv^2$, and takes the form~\cite{Alshits:1979,Brailsford:1972}
\begin{align}
 &B=\frac{4\pi}{\hbar v^2}\sum_{s',s''}\sum\limits_{q',q''}\int\!\!d^2q\, \W_q |\G_{s's''}(\vqi,\vqii,\vq)|^2(n_{q''}-n_{q'})\nn\\
 &\quad\times \d(\vqi-\vqii-\vq)\d(\w_{q'}-\w_{q''}-\W_q)
 \,,\nn\\
 &\G_{s's''}(\vqi,\vqii,\vq)= \hbar\sum_{i,j,k}\frac{ d_{kk'}(\vq) \mathbf{w}_{q'i} ^*\mathbf{w}_{q''j} }{4\rho \sqrt{\w_{q'}\w_{q''}}}
 \sum_{i'j'k'} q'_{i'}q''_{j'} \tilde{A}_{ijk}^{i'j'k'}
 \,, \label{eq:dissipation-alshits1979}
\end{align}
where $\vqii$, $\vqi$ are the wave vectors of incoming and outgoing phonons, $s'$, $s''$ label their polarizations (2 transverse, 1 longitudinal), and $\vq$ is the wave vector associated with the dislocation field in Fourier space, $d_{kk'}$.
In contrast to the phonons (which are quantized and have discrete wave vectors determined from the perfect lattice), the dislocation is modeled as a classical field in the continuum limit.
Assuming an infinitely long, straight dislocation, its only spatial dependence is within the plane perpendicular to the dislocation line.
The sums over discrete phonon momenta can subsequently be approximated as integrals over the first Brillouin zone.
$\w_{q'}$, $\w_{q''}$ are the phonon frequencies presently depending linearly on the wave vector length in accordance with the isotropic Debye approximation, i.e. $\w_{q'}=c_{s'}\abs{\vqi}$ where $c_{s'}$ is the sound speed of a phonon with polarization $s'$ (either transverse $\ct$ or longitudinal $\cl$).
$\W_q=\abs{\vq\cdot\vv{v}}$ is the energy transfer whenever a phonon scatters on the dislocation.
$n_{q'}$ denotes the equilibrium phonon distribution function $n_{q'}=(\exp(\hbar\w_{q'}/k_BT)-1)^{-1}$, which controls the number of scattering events per unit time.
$\hbar$ is Planck's constant, $\rho$ is the material density, and the two Dirac delta functions in the second line of \eqref{eq:dissipation-alshits1979} encode momentum and energy conservation within each scattering event.
$\G$ finally represents the associated matrix element, or scattering probability.
As such, it depends on the (anisotropic) dislocation displacement gradient field $d_{kk'}$, the (quantized, isotropic) phonons whose orthonormal polarization vectors are presently denoted by $\bfw_{qi}\coleq\bfw_i(\vq,s)$, and a linear combination of second (SOEC) and third order elastic constants (TOEC) of the anisotropic single crystal grains of a polycrystal, $\tilde{A}_{ijk}^{i'j'k'}$.
For technical details on the theory we refer to~\cite{Blaschke:2018anis} as well as~\cite{Alshits:1992,Blaschke:BpaperRpt}.

\subsection{Steady state dislocations and slip geometries}
\label{sec:disloc+slips}

\begin{table*}[h!t!b]
{\renewcommand{\arraystretch}{1.2}
\small
\centering
 \begin{tabular}{c|c|c|c|c|c|c|c|c}
 & Ag$_\txt{(fcc)}$ & Al$_\txt{(fcc)}$ & Au$_\txt{(fcc)}$ & Cu$_\txt{(fcc)}$ & Fe$_\txt{(bcc)}$ & Mo$_\txt{(bcc)}$ & Nb$_\txt{(bcc)}$ & Ni$_\txt{(fcc)}$ \\\hline\hline
 $a$[\r{A}] & 4.09 & 4.05 & 4.08 & 3.61 & 2.87 & 3.15 & 3.30 & 3.52\\
 $\rho$[g/ccm] & 10.50 & 2.70 & 19.30 & 8.96 & 7.87 & 10.20 & 8.57 & 8.90\\
 $\lambda$[GPa] & 83.6 & 58.1 & 198.0 & 105.5 & 115.5 & 176.4$^*$ & 144.5 & 126.1\\
 $\mu$[GPa] & 30.3 & 26.1 & 27.0 & 48.3 & 81.6 & 125.0$^*$ & 37.5 & 76.0
\\\hline\hline
 $c_{11}$ & 123.99 & 106.75 & 192.44 & 168.30 & 226.00 & 463.70 & 246.50 & 248.10\\
 $c_{12}$ & 93.67 & 60.41 & 162.98 & 121.20 & 140.00 & 157.80 & 134.50 & 154.90\\
 $c_{44}$ & 46.12 & 28.34 & 42.00 & 75.70 & 116.00 & 109.20 & 28.73 & 124.20
\\\hline\hline
 $A$ & 3.04 & 1.22 & 2.85 & 3.21 & 2.70 & 0.71 & 0.51 & 2.67
\\\hline\hline
 $c_{111}$ & $-843$ & $-1076$ & $-1729$ & $-1271$ & $-2720$ & $-3557$ & $-2564$ & $-2040$\\
 $c_{112}$ & $-529$ & $-315$ & $-922$ & $-814$ & $-608$ & $-1333$ & $-1140$ & $-1030$\\
 $c_{123}$ & $189$ & $36$ & $-233$ & $-50$ & $-578$ & $-617$ & $-467$ & $-210$\\
 $c_{144}$ & $56$ & $-23$ & $-13$ & $-3$ & $-836$ & $-269$ & $-343$ & $-140$\\
 $c_{166}$ & $-637$ & $-340$ & $-648$ & $-780$ & $-530$ & $-893$ & $-168$ & $-920$\\
 $c_{456}$ & $83$ & $-30$ & $-12$ & $-95$ & $-720$ & $-555$ & $137$ & $-70$
\end{tabular}
\caption{%
List of input data for cubic crystals used in the calculation of the drag coefficient; all elastic constants are given in units of GPa.
The references we used to compile these data are: Ref.~\cite[Sec. 12]{CRCHandbook} (lattice parameters $a$ and densities $\rh$),
Refs.~\cite[p.~10]{Hertzberg:2012} and~\cite{Kaye:2004} (effective Lam\'e constants of the polycrystal except for Mo),
Ref.~\cite[Sec. 12]{CRCHandbook} (single crystal SOEC and Zener anisotropy ratio $A\coleq 2c_{44}/(c_{11}-c_{12})$),
and Refs.~\cite{Thomas:1968,Hiki:1966,Powell:1984,Voronov:1978,Graham:1968,Riley:1973} (TOEC).
The Lam\'e constants of Mo (marked with $^*$) are analytical averages of the single crystal SOEC, see e.g.~\cite{Blaschke:2017Poly}.
The conventions for the single crystal elastic constants are those of Brugger~\cite{Brugger:1965}.
}
\label{tab:data_cubic}
}
\end{table*}

\begin{table}[h!t!b]
{\renewcommand{\arraystretch}{1.2}
\small
\centering
 \begin{tabular}{c|c|c|c|c|c}
 (hcp) & Cd & Mg & Ti & Zn & Zr \\\hline\hline
 $a$[\r{A}] & 2.98 & 3.21 & 2.95 & 2.67 & 3.23\\
 $c$[\r{A}] & 5.62 & 5.21 & 4.68 & 4.95 & 5.15\\
 $\rho$[g/ccm] & 8.69 & 1.74 & 4.51 & 7.13 & 6.52\\
 $\lambda$[GPa] & 28.8 & 24.1 & 78.5 & 43.1 & 71.3$^*$\\
 $\mu$[GPa] & 19.2 & 17.3 & 43.8 & 43.4 & 36.0$^*$
\\\hline\hline
 $c_{11}$ & 114.50 & 59.50 & 162.40 & 163.68 & 143.40\\
 $c_{12}$ & 39.50 & 26.12 & 92.00 & 36.40 & 72.80\\
 $c_{44}$ & 19.85 & 16.35 & 46.70 & 38.79 & 32.00\\
 $c_{13}$ & 39.90 & 21.80 & 69.00 & 53.00 & 65.30\\
 $c_{33}$ & 50.85 & 61.55 & 180.70 & 63.47 & 164.80
\\\hline\hline
 $c_{111}$ & $-2060$ & $-663$ & $-1358$ & $-1760$ & $-767$\\
 $c_{112}$ & $-114$ & $-178$ & $-1105$ & $-440$ & $-697$\\
 $c_{123}$ & $-110$ & $-76$ & $-162$ & $-210$ & $37$\\
 $c_{144}$ & $227$ & $-30$ & $-263$ & $-10$ & $37$
\\\hline
 $c_{113}$ & $-197$ & $30$ & $17$ & $-270$ & $-96$\\
 $c_{133}$ & $-268$ & $-86$ & $-383$ & $-350$ & $-271$\\
 $c_{155}$ & $-332$ & $-58$ & $117$ & $250$ & $-271$\\
 $c_{222}$ & $-2020$ & $-864$ & $-2306$ & $-2410$ & $-1450$\\
 $c_{333}$ & $-516$ & $-726$ & $-1617$ & $-720$ & $-2154$\\
 $c_{344}$ & $-171$ & $-193$ & $-383$ & $-440$ & $-271$
\end{tabular}
\caption{%
List of input data for hcp crystals used in the calculation of the drag coefficient; all elastic constants are given in units of GPa.
The references we used to compile these data are: Ref.~\cite[Sec. 12]{CRCHandbook} (lattice parameters $a$, $c$ and densities $\rh$),
Refs.~\cite[p.~10]{Hertzberg:2012} and~\cite{Kaye:2004} (effective Lam\'e constants of the polycrystal except for Zr),
Ref.~\cite[Sec. 12]{CRCHandbook} (single crystal SOEC),
and Refs.~\cite{Saunders:1986,Naimon:1971,Rao:1973,Swartz:1970,Singh:1992} (TOEC).
The Lam\'e constants of Zr (marked with $^*$) are analytical averages of the single crystal SOEC, see e.g.~\cite{Blaschke:2017Poly}.
The conventions for the single crystal elastic constants are those of Brugger~\cite{Brugger:1965}.
}
\label{tab:data_hcp}
}
\end{table}

The displacement gradient field in the continuum limit and within the realm of linear elasticity of a dislocation moving at constant velocity, can be determined from solving the equations of motion (e.o.m.) and the (leading order) stress-strain relations known as Hooke's law:
\begin{align}
 \pa_i\s_{ij}&=\rh\ddot u_j\,, &
 \s_{ij}&=C_{ijkl}\e_{kl}=C_{ijkl}u_{k,l}
 \,, \label{eq:Hooke}
\end{align}
where we have introduced the notation
$u_{k,l}\coleq\pa_l u_k$ for the gradient of the displacement field $u_k$, and $\ddot u_j\coleq \frac{\pa^2 u_j}{\pa t^2}$ for the time derivatives.
For constant velocity $v_i$, this system of equations can be rewritten as $\hat C_{ijkl}u_{k,il}=0$ with ``effective'' elastic constants $\hat C_{ijkl}\coleq\left(C_{ijkl}-\rho v_iv_l \d_{jk}\right)$, see~\cite{Bacon:1980}.
Upon introducing perpendicular unit vectors $\vec{m}_0$ and $\vec{n}_0$ which are normal to the sense vector $\vec{t}$ of the dislocation, i.e. $\vec{t}=\vec{m}_0\times\vec{n}_0$,
the solution takes the form $u_{j,k}(r,\phi)={\tilde{u}_{j,k}(\phi)}/{r}$ where $\tilde{u}_{j,k}(\phi)$ is a function of Burgers vector, $\vec{m}$, $\vec{n}$, and $\hat C_{ijkl}$~\cite[p.~476]{Hirth:1982}:
\begin{align}
 \tilde{u}_{\!j,k}&=\frac{b_l}{2\pi}\left\{n_k\left[(nn)^{-1}(nm)\!\cdot\!S\right]_{jl} - m_k S_{jl} + n_k(nn)^{-1}_{ji}K_{il}\right\}
 , \nn\\
 \mat S&=-\frac1{2\pi}\int_0^{2\pi}(nn)^{-1}(nm)\,d\phi
 \,, \nn\\
 \mat K&=-\frac1{2\pi}\int_0^{2\pi}\left[(mn)(nn)^{-1}(nm)-(mm)\right]d\phi
 \,,\label{eq:ukl-sol}
\end{align}
with the shorthand notation $(ab)_{jk}\coleq a_i \hat C_{ijkl} b_l$.
Variables $r$, $\phi$ are polar coordinates in the plane spanned by $\vec{m}=\vec{m}_0(\vth)\cos\ph + \vec{n}_0\sin\ph$ and $\vec{n}=\vec{n}_0\cos\ph - \vec{m}_0(\vth)\sin\ph$, where $\vec{n}_0$ is the slip plane normal and $\vec{m}_0(\vth)$ is perpendicular to $\vec{n}_0$ and $\vec{t}(\vth) = \frac1b\left[\vec{b}\cos\vth+\vec{b}\times\vec{n}_0\sin\vth\right]$.
As such, $\vec{m}_0(\vth)$ depends on the dislocation character angle $\vth$ and is parallel to $\vec{v}$.
The important feature to note is that $\tilde{u}_{j,k}(\phi)$ includes terms proportional to
$(nn)^{-1}$ and hence exhibits divergences whenever $\det(nn)=0$.
This happens at certain combinations of polar angle $\phi$ and critical velocity $\abs{\vec{v}_\txt{c}}$.
As was shown in Ref.~\cite{Blaschke:2017lten,Blaschke:2021vcrit}, critical velocities are typically close to (and sometimes equal to) the lowest shear wave speed associated with the direction of $\vv{v}$ in the single crystal.
All dislocation displacement gradients computed with the present method are hence restricted to (constant) velocities $v$ which are \emph{smaller} than $\vc$.

As noted in the previous section, dislocation field $u_{j,k}(r,\phi)$ (more precisely its Fourier transform $d_{jk}(\vq)$) enters $\G$ within \eqref{eq:dissipation-alshits1979}, and thus the drag coefficient $B$ depends quadratically on $u_{j,k}$.
For simplicity, we presently only consider perfect dislocations;
Incorporating more realistic models of the dislocation core as well as the effect of partial dislocations into the dislocation drag coefficient are beyond the scope of the present paper and we leave those considerations to future work.
For recent advances on the theoretical modeling of dislocation cores (albeit disconnected from phonon wind theory), see~\cite{Clouet:2011a,Szajewski:2017phil,Pellegrini:2018,Boleininger:2018,Blaschke:2019core} and references therein.

The slip systems we have considered here are:
\begin{rstrip}
\begin{align}
 &\vec{b}^\txt{fcc}=\frac{b^\txt{fcc}}{\sqrt{2}}\left(1,1,0\right)
 \,, &&
 b^\txt{fcc}=\frac{a}{\sqrt{2}}
 \,, &&
 \vec{n}_0^\txt{fcc}=\frac{1}{\sqrt{3}}\left(-1,1,-1\right)
 \,, \nn\\
 &\vec{b}^\txt{bcc}=\frac{b^\txt{bcc}}{\sqrt{3}}\left(1,-1,1\right)
 \,, &&
 b^\txt{bcc}=\frac{a\sqrt{3}}{2}
 \,, &&
 \vec{n}_0^\txt{bcc}=\frac{1}{\sqrt{2}}\left(1,1,0\right)
 \,, \nn\\
 &\vec{b}^\txt{hcp}=b^\txt{hcp}\left(-1,0,0\right)
 \,, &&
 b^\txt{hcp}=a
 \,, &&
 \vec{n}_0^\txt{hcp-basal}=\left(0,0,1\right)
 \,, \nn\\
 &\vec{n}_0^\txt{hcp-prismatic}=\left(0,-1,0\right)
 \,, &&
 \vec{n}_0^\txt{hcp-pyramidal}=\frac1{\sqrt{\frac34a^2+c^2}}\left(0,-c,\frac{\sqrt3}{2}a\right)
 \,, \label{eq:slipplanes}
\end{align}
\end{rstrip}
\noindent
where $a$ and $c$ are the lattice constants given in Tables~\ref{tab:data_cubic} and~\ref{tab:data_hcp},
and $\vec{b}$, $\vec{n}_0$ denote the Burgers vector and slip plane normals, respectively;
see Refs.~\cite{Blaschke:2017lten,Blaschke:2018anis} for details.
For the case of close-packed hexagonal (hcp) crystals, we assume the basal plane is normal to the third axis in Cartesian crystal coordinates.
The three hcp slip systems we consider, basal, prismatic, and pyramidal slip, share the same Burgers vector but have different slip plane normals.
All except for the bcc slip system above lead to expressions which are symmetric with respect to $\vth\to-\vth$, and all slip systems are $\pi$-periodic.

In Tables~\ref{tab:data_cubic} and~\ref{tab:data_hcp} we list all input data that were used in the computation of the drag coefficient below.
For the effective Lam\'e constants of the polycrystal, we have chosen to use the separate experimental values (where available) listed in those tables rather than analytically averaging over the single crystal values.
The only exceptions are Mo and Zr due to lack of experimental data, and because the Voigt and Reuss bounds are very close to each other in those cases.
In fact, for SOEC of cubic crystals, analytic averaging would be a viable avenue as well\footnote{
In fact, the single crystal averages for the Lam\'e constants of cubic crystals agree well (within a few percent) with the experimental results listed in Table~\ref{tab:data_cubic},
with the exception of Ni whose averaged shear modulus is $\sim11\%$ higher than the measured value,
and also Au whose averaged $\l$ is $\sim12\%$ lower than the measured value.
}
(assuming negligible texturing), but not so much for hcp and other crystals; see~\cite{Blaschke:2017Poly} and references therein.

\subsection{The low velocity limit}
\label{sec:lowvelocity}

In the limit of small velocity $v$, small meaning $v\ll\ct$ and $v\ll \cs$ (where $\ct$ is the effective polycrystalline transverse sound speed and $\cs$ is the lowest shear wave speed of the single crystal in the direction of $\vv{v}$),
drag coefficient $B$ simplifies to
\begin{align}
 B&\approx
 \frac{4\pi}{\hbar }\sum_{s',s''}\int_\txt{BZ}\!\frac{d^3q'}{(2\pi)^3}\int_\txt{BZ}\!\frac{d^2q}{(2\pi)^2}\;
 \abs{\Gamma_{s's''}(\vqi,\vqi\!\!-\vq,\vq,v=0)}^2
 \nn\\&\quad\times
  (\vq\cdot\hat{v})^2 \frac{\partial (-n_{q'})}{\partial\omega_{q'}}  \delta(\omega_{q'}-\omega_{q'-q}) 
 + C\, v + \cO(v^2)
 \,, \label{eq:Bsmallv}
\end{align}
where $\hat{v}$ denotes the unit vector in the direction of $\vv{v}$.
Explicit numerical calculations for a number of metals show that the first order velocity correction has a \emph{negative} coefficient $C<0$.
To understand why this is the case, we note that the dislocation field itself depends only on the square of its velocity and thus its Taylor expansion around small $v$ has no linear term.
Furthermore, since $\w_{q'}=c_{s'}q'$ and $\G$ scales like $1/\w_{q'}\w_{q''}$, the drag coefficient depends on the sound speeds as $1/c_{s'}^3c_{s''}^2$.
Since $\ct\sim\cl/2$, the largest contribution to $B$ at low velocity $v$ is due to the purely transverse branch (where both incoming and outgoing phonons are transverse),
as was already observed in earlier work~\cite{Alshits:1979,Blaschke:BpaperRpt,Blaschke:2018anis}.
In this case, it is convenient to introduce a dimensionless integration variable proportional to the ratio $t\propto\abs{\vqi}/\abs{\vq}$.
The energy conserving delta function then restricts the integration range of this new variable $t$ such that it shrinks with growing dislocation velocity $v$, see~\cite{Blaschke:BpaperRpt,Blaschke:2018anis}.
This is the dominating effect and the reason for negative $C$.


\begin{figure*}[h!t!b]
 \centering
 \includegraphics[page=1,width=0.368\textwidth,trim={0.5mm 2mm 2mm 2mm},clip]{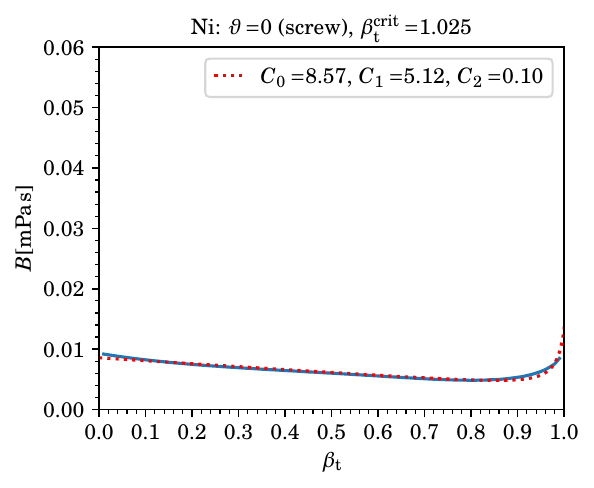}%
 \includegraphics[page=2,width=0.316\textwidth,trim={1.45cm 2mm 2mm 2mm},clip]{B_fit7_Ni_3c.pdf}%
 \includegraphics[page=3,width=0.316\textwidth,trim={1.45cm 2mm 2mm 2mm},clip]{B_fit7_Ni_3c.pdf}
 \includegraphics[page=4,width=0.368\textwidth,trim={0.5mm 2mm 2mm 2mm},clip]{B_fit7_Ni_3c.pdf}%
 \includegraphics[page=5,width=0.316\textwidth,trim={1.45cm 2mm 2mm 2mm},clip]{B_fit7_Ni_3c.pdf}%
 \includegraphics[page=6,width=0.316\textwidth,trim={1.45cm 2mm 2mm 2mm},clip]{B_fit7_Ni_3c.pdf}
 \includegraphics[page=7,width=0.368\textwidth,trim={0.5mm 2mm 2mm 2mm},clip]{B_fit7_Ni_3c.pdf}\vspace{-0.25cm}
\caption{%
We show the drag coefficient $B(\bt)$ from phonon wind for dislocations in Ni of various character angles $\vth$.
The dashed lines represent the three-parameter fitting functions with fitting parameters $C_i$ in units of $\mu$Pa$\,$s, critical velocity $\beta_\mathrm{t}^\mathrm{crit}=\vc/\ct$, and $\bt=v/\ct$.
The solid lines show the results of numerically evaluating $B$ according to \eqnref{eq:dissipation-alshits1979} using the software of Ref.~\cite{pydislocdyn}, see Ref.~\cite{Blaschke:2018anis} for details on the method.}
 \label{fig:drag-Ni-theta}
\end{figure*}

\subsection{High velocity limit}


Our use of an isotropic Debye phonon spectrum introduces the limitation $v<\ct$ on our present theory.
Nonetheless, $B$ does not diverge at $v=\ct$:
All divergences within $B$ are inherited from the poles present in the dislocation field, as pointed out in Sec.~\ref{sec:disloc+slips} above.
Indeed, those appear at critical velocities $\vc$ which depend on the slip geometry, material constants, as well as the dislocation character $\vth$.
In order to determine the highest degree of divergence, we first recall the study done in Ref.~\cite{Blaschke:BpaperRpt} in the purely isotropic limit and only for the transverse phonon modes:
There it was found that the highest degree of divergence of a dislocation field for pure edge was $1/(1-\bt^2)^m$ with $\bt\coleq v/\ct$ and $m=1$ at polar angle $\phi=0$ (or $\pi$), whereas the one for pure screw exhibited the milder divergence of $m=1/2$.
Within $B$, where the dislocation field enters quadratically and angles $\phi$ are integrated over, this led to initial estimates for the degree of divergence of $B$ of $m=3/2$ for pure edge and $m=1/2$ for pure screw.
However, within the purely transverse branch, the kinematic terms in $\G$ additionally suppressed the degree of divergence by 1, ultimately leading to $B\sim 1/(1-\bt^2)^m$ as $\bt\to1$ with $m=1/2$ for edge and finite $B$ for screw dislocations.

In the more general semi-isotropic case considered here, this latter cancellation cannot occur because now we have divergences at $\vc(\vth)$ whereas the kinematic terms in $\G$ coming from the phonons only know about $\ct$, $\cl$.
Likewise, the cancellation leading to the milder divergence of the pure screw dislocation in the isotropic limit is indeed special to the strictly isotropic case:
For an isotropic screw dislocation, $\mat{S}\!\cdot\!\vv{b}\to\vv0$, $\mat{K}\!\cdot\!\vv{b}\sim\left(0,0,\sqrt{1-\bt^2}\right)$, and $(nn)^{-1}\sim 1/(1-\bt^2)$ within \eqref{eq:ukl-sol} yield the milder divergence noted above.
Finally, one must also not forget that edge and screw dislocations  decouple only in the isotropic limit, but not in general, which is why mixed dislocations cannot be represented as superpositions of edge and screw in ``real'' crystals.

To sum up: we presently expect the highest degree of divergence of the drag coefficient $B(v,\vth)$ at $v\to\vc(\vth)$ to be $1/(1-v^2/\vc^2)^m$ with $m=3/2$ for arbitrary dislocation characters $\vth$.
Indeed, this expectation is confirmed by numerical results, where the asymptotic region cannot be well represented by fitting functions with $m<3/2$.

\section{Results and their effective functional form}


\begin{figure*}[h!t!b]
 \centering
 \includegraphics[width=0.368\textwidth,trim={1mm 2mm 2mm 2mm},clip]{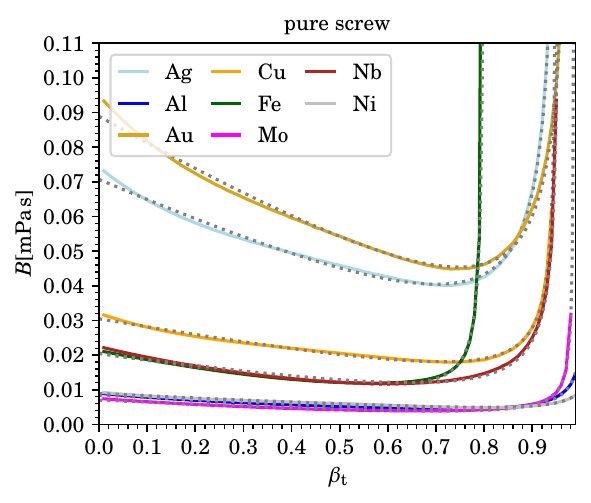}%
 \includegraphics[width=0.316\textwidth,trim={1.45cm 2mm 2mm 2mm},clip]{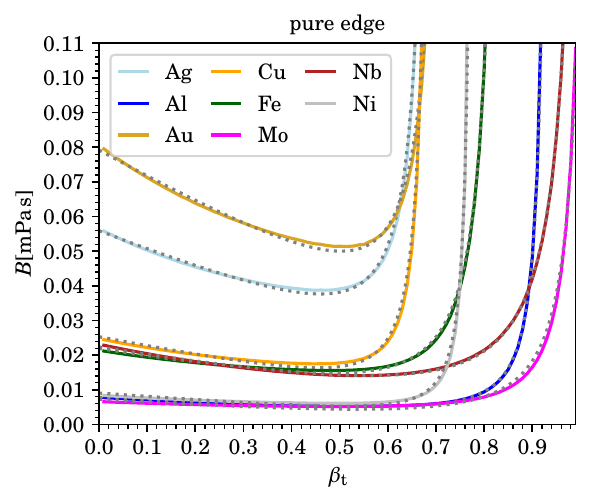}%
 \includegraphics[width=0.316\textwidth,trim={1.45cm 2mm 2mm 2mm},clip]{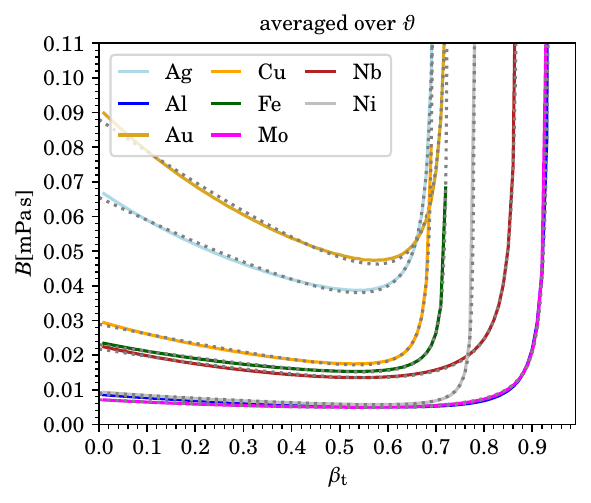}\vspace*{-0.25cm}
\caption{%
We show the drag coefficient $B(\bt)$ from phonon wind for pure screw and edge dislocations as well as $B$ averaged over all character angles $\vth$ for 5 fcc and 3 bcc metals.
The dashed lines represent the fitting functions and $\bt=v/\ct$.}
 \label{fig:drag-cub}
\end{figure*}

\begin{table*}[h!t!b]
{\renewcommand{\arraystretch}{1.2}
\small
\centering
 \begin{tabular}{c|c|c|c|c|c|c|c|c}
  & Ag$_\txt{(fcc)}$ & Al$_\txt{(fcc)}$ & Au$_\txt{(fcc)}$ & Cu$_\txt{(fcc)}$ & Fe$_\txt{(bcc)}$ & Mo$_\txt{(bcc)}$ & Nb$_\txt{(bcc)}$ & Ni$_\txt{(fcc)}$ \\\hline\hline
 $\ct$ & 1699 & 3109 & 1183 & 2322 & 3220 & 3501 & 2092 & 2922 \\
 $v_c^{\mathrm{e}}/\ct$ & 0.707 & 0.942 & 0.739 & 0.698 & 0.852 & 1.033 & 1.026 & 0.783 \\
 $v_c^{\mathrm{e}}$ & 1202 & 2929 & 874 & 1621 & 2745 & 3615 & 2147 & 2288 \\
 $C_0^{\mathrm{e}}$ & $55.89$ & $7.85$ & $79.16$ & $25.30$ & $21.92$ & $7.19$ & $22.36$ & $9.06$ \\
 $C_1^{\mathrm{e}}$ & $37.62$ & $6.56$ & $56.07$ & $18.49$ & $18.49$ & $8.26$ & $22.68$ & $7.81$ \\
 $C_2^{\mathrm{e}}$ & $0.00$ & $0.00$ & $0.00$ & $0.00$ & $0.00$ & $0.00$ & $6.67$ & $0.00$ \\
 $C_3^{\mathrm{e}}$ & $4.85$ & $1.25$ & $5.83$ & $2.38$ & $4.23$ & $2.45$ & $4.07$ & $1.01$
\\\hline\hline
 $v_c^{\mathrm{s}}/\ct$ & 0.952 & 1.005 & 0.981 & 0.950 & 0.803 & 0.987 & 0.955 & 1.025 \\
 $v_c^{\mathrm{s}}$ & 1617 & 3126 & 1160 & 2205 & 2585 & 3457 & 1997 & 2996 \\
 $C_0^{\mathrm{s}}$ & $70.72$ & $8.43$ & $88.92$ & $30.49$ & $20.42$ & $7.01$ & $20.56$ & $8.89$ \\
 $C_1^{\mathrm{s}}$ & $56.61$ & $7.28$ & $76.72$ & $22.52$ & $15.96$ & $6.15$ & $18.08$ & $6.71$ \\
 $C_2^{\mathrm{s}}$ & $23.88$ & $2.96$ & $27.48$ & $8.75$ & $5.45$ & $2.76$ & $10.38$ & $2.11$ \\
 $C_3^{\mathrm{s}}$ & $0.00$ & $0.00$ & $0.00$ & $0.00$ & $0.14$ & $0.02$ & $0.00$ & $0.00$
\\\hline\hline
 $v_c^{\mathrm{av}}/\ct$ & 0.707 & 0.942 & 0.739 & 0.698 & 0.726 & 0.935 & 0.875 & 0.783 \\
 $v_c^{\mathrm{av}}$ & 1202 & 2929 & 874 & 1621 & 2337 & 3272 & 1831 & 2288 \\
 $C_0^{\mathrm{av}}$ & $65.49$ & $8.72$ & $88.00$ & $28.91$ & $23.00$ & $7.10$ & $21.77$ & $9.13$ \\
 $C_1^{\mathrm{av}}$ & $47.73$ & $9.02$ & $69.82$ & $20.25$ & $14.37$ & $6.12$ & $19.28$ & $6.37$ \\
 $C_2^{\mathrm{av}}$ & $14.79$ & $5.46$ & $19.37$ & $6.14$ & $5.42$ & $5.36$ & $13.12$ & $2.23$ \\
 $C_3^{\mathrm{av}}$ & $0.32$ & $0.22$ & $0.37$ & $0.14$ & $0.05$ & $0.10$ & $0.09$ & $0.07$
\end{tabular}
\caption{List of critical velocities $\vc$[m/s], and fitting function coefficients $C_i$[$\mu$Pa\,s] for some cubic crystals.
The critical velocities are given in units of m/s as well as in ratio to $\ct$.
The fits are only valid up to $0.99\ct$ and do not capture the asymptotic behavior for those metals/dislocations whose critical velocity is larger than this value.
Superscripts ``e'', ``s'', and ``av'' refer to ``edge'', ``screw'', and ``average'', respectively.
Furthermore, $\vc^{\text{av}}$ coincides with the smallest critical velocity for all dislocation characters $\vth$ within the slip system.}
\label{tab:coefficients_cub}
}
\end{table*}


\begin{figure*}[h!t!b]
 \centering
 \includegraphics[width=0.368\textwidth,trim={1mm 2mm 2mm 2mm},clip]{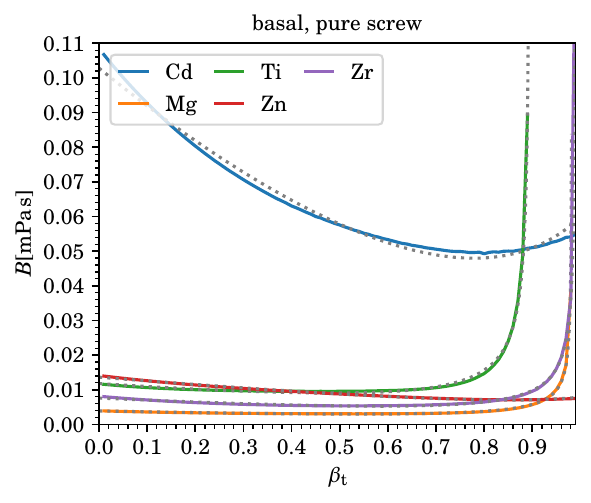}%
 \includegraphics[width=0.316\textwidth,trim={1.45cm 2mm 2mm 2mm},clip]{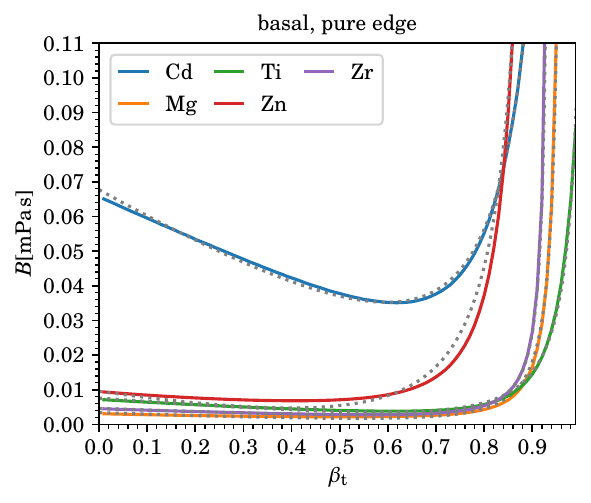}%
 \includegraphics[width=0.316\textwidth,trim={1.45cm 2mm 2mm 2mm},clip]{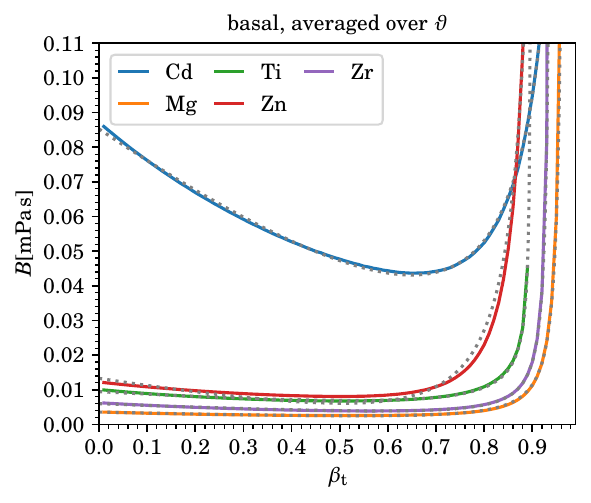}\vspace{-0.25cm}
\caption{%
We show the drag coefficient $B(\bt)$ from phonon wind for pure screw and edge dislocations as well as averaged over all character angles $\vth$ for basal slip of 5 hcp metals.
Dashed lines represent the fitting functions and $\bt=v/\ct$.}
 \label{fig:drag-bas}
\end{figure*}

\begin{table}[h!t!b]
{\renewcommand{\arraystretch}{1.2}
\small
\centering
 \begin{tabular}{c|c|c|c|c|c}
 basal & Cd & Mg & Ti & Zn & Zr \\\hline\hline
 $\ct$ & 1486 & 3153 & 3118 & 2466 & 2350 \\
 $v_c^{\mathrm{e}}/\ct$ & 1.017 & 0.972 & 1.033 & 0.943 & 0.943 \\
 $v_c^{\mathrm{e}}$ & 1511 & 3065 & 3219 & 2326 & 2215 \\
 $C_0^{\mathrm{e}}$ & $67.73$ & $3.37$ & $7.59$ & $9.56$ & $4.61$ \\
 $C_1^{\mathrm{e}}$ & $78.74$ & $4.42$ & $10.40$ & $18.73$ & $4.62$ \\
 $C_2^{\mathrm{e}}$ & $0.00$ & $0.00$ & $1.17$ & $0.00$ & $1.25$ \\
 $C_3^{\mathrm{e}}$ & $15.53$ & $1.08$ & $2.05$ & $9.05$ & $0.64$
\\\hline\hline
 $v_c^{\mathrm{s}}/\ct$ & 1.398 & 0.982 & 0.896 & 1.211 & 0.990 \\
 $v_c^{\mathrm{s}}$ & 2077 & 3097 & 2795 & 2987 & 2327 \\
 $C_0^{\mathrm{s}}$ & $102.80$ & $3.88$ & $11.79$ & $13.62$ & $7.60$ \\
 $C_1^{\mathrm{s}}$ & $157.93$ & $2.69$ & $8.22$ & $13.32$ & $6.25$ \\
 $C_2^{\mathrm{s}}$ & $162.34$ & $2.89$ & $8.93$ & $6.81$ & $5.58$ \\
 $C_3^{\mathrm{s}}$ & $0.00$ & $0.00$ & $0.04$ & $0.00$ & $0.01$
\\\hline\hline
 $v_c^{\mathrm{av}}/\ct$ & 1.017 & 0.972 & 0.896 & 0.943 & 0.943 \\
 $v_c^{\mathrm{av}}$ & 1511 & 3065 & 2795 & 2326 & 2215 \\
 $C_0^{\mathrm{av}}$ & $85.26$ & $3.55$ & $9.45$ & $13.28$ & $5.97$ \\
 $C_1^{\mathrm{av}}$ & $95.45$ & $2.91$ & $6.70$ & $20.08$ & $4.98$ \\
 $C_2^{\mathrm{av}}$ & $42.76$ & $0.66$ & $5.90$ & $0.00$ & $2.40$ \\
 $C_3^{\mathrm{av}}$ & $5.03$ & $0.54$ & $0.00$ & $5.39$ & $0.31$
\end{tabular}
\caption{List of critical velocities $\vc$[m/s], and fitting function coefficients $C_i$[$\mu$Pa\,s] for basal slip.
The critical velocities are given in units of m/s as well as in ratio to $\ct$.
The fits are only valid up to $0.99\ct$ and do not capture the asymptotic behavior for those metals/dislocations whose critical velocity is larger than this value.
Superscripts ``e'', ``s'', and ``av'' refer to ``edge'', ``screw'', and ``average'', respectively.
Furthermore, $\vc^{\text{av}}$ coincides with the smallest critical velocity for all dislocation characters $\vth$ within the basal slip system.}
\label{tab:coefficients_bas}
}
\end{table}


\begin{figure*}[h!t!b]
 \centering
 \includegraphics[width=0.368\textwidth,trim={1mm 2mm 2mm 2mm},clip]{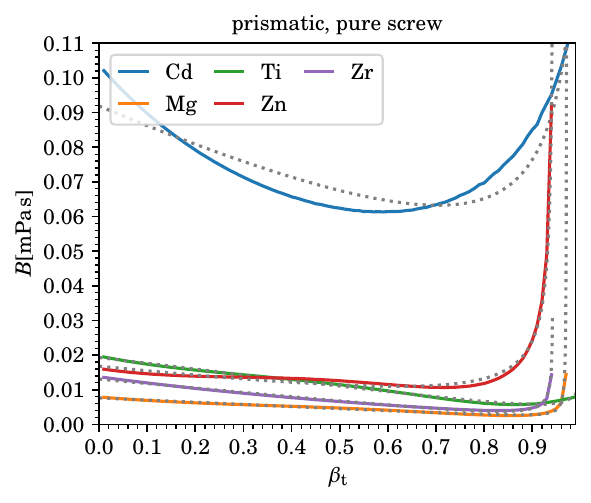}%
 \includegraphics[width=0.316\textwidth,trim={1.45cm 2mm 2mm 2mm},clip]{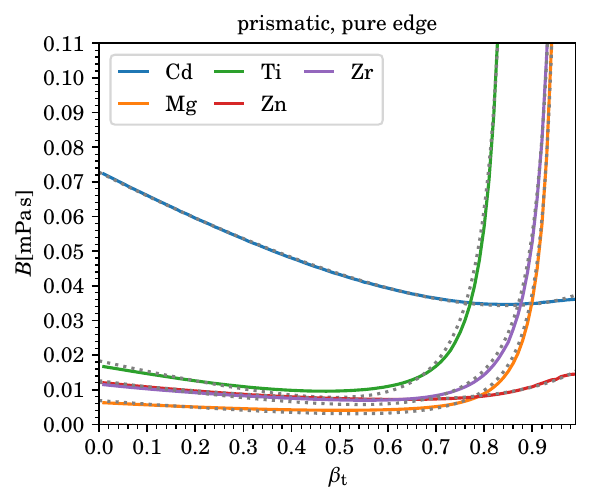}%
 \includegraphics[width=0.316\textwidth,trim={1.45cm 2mm 2mm 2mm},clip]{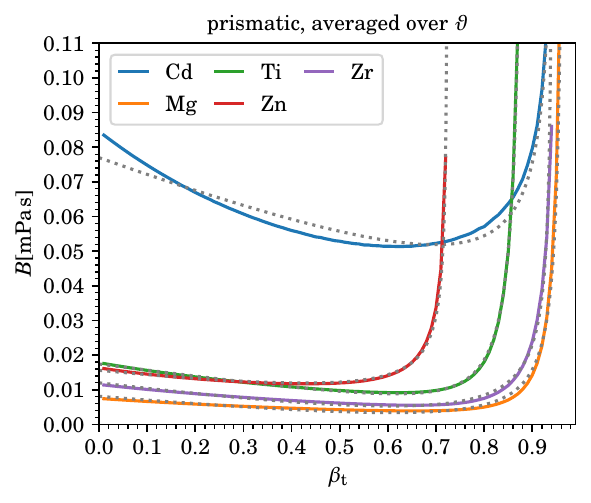}\vspace{-0.25cm}
\caption{%
We show the drag coefficient $B(\bt)$ from phonon wind for pure screw and edge dislocations as well as averaged over all character angles $\vth$ for prismatic slip of 5 hcp metals.
Dashed lines represent the fitting functions and $\bt=v/\ct$.}
 \label{fig:drag-pris}
\end{figure*}

\begin{table}[h!t!b]
{\renewcommand{\arraystretch}{1.2}
\small
\centering
 \begin{tabular}{c|c|c|c|c|c}
 prismatic & Cd & Mg & Ti & Zn & Zr \\\hline\hline
 $\ct$ & 1486 & 3153 & 3118 & 2466 & 2350 \\
 $v_c^{\mathrm{e}}/\ct$ & 1.398 & 0.982 & 0.896 & 1.211 & 0.990 \\
 $v_c^{\mathrm{e}}$ & 2077 & 3097 & 2795 & 2987 & 2327 \\
 $C_0^{\mathrm{e}}$ & $72.77$ & $6.88$ & $18.35$ & $12.13$ & $12.61$ \\
 $C_1^{\mathrm{e}}$ & $98.31$ & $10.56$ & $28.29$ & $16.52$ & $18.57$ \\
 $C_2^{\mathrm{e}}$ & $82.02$ & $0.00$ & $0.00$ & $18.90$ & $0.00$ \\
 $C_3^{\mathrm{e}}$ & $0.00$ & $2.80$ & $7.00$ & $0.66$ & $4.65$
\\\hline\hline
 $v_c^{\mathrm{s}}/\ct$ & 1.017 & 0.972 & 1.033 & 0.945 & 0.943 \\
 $v_c^{\mathrm{s}}$ & 1511 & 3065 & 3219 & 2332 & 2215 \\
 $C_0^{\mathrm{s}}$ & $91.85$ & $7.65$ & $19.30$ & $16.80$ & $13.09$ \\
 $C_1^{\mathrm{s}}$ & $59.31$ & $6.21$ & $17.41$ & $13.04$ & $12.40$ \\
 $C_2^{\mathrm{s}}$ & $32.21$ & $0.44$ & $0.95$ & $8.37$ & $1.42$ \\
 $C_3^{\mathrm{s}}$ & $0.00$ & $0.01$ & $0.09$ & $0.02$ & $0.00$
\\\hline\hline
 $v_c^{\mathrm{av}}/\ct$ & 0.948 & 0.972 & 0.896 & 0.724 & 0.943 \\
 $v_c^{\mathrm{av}}$ & 1409 & 3065 & 2795 & 1786 & 2215 \\
 $C_0^{\mathrm{av}}$ & $76.98$ & $8.10$ & $17.44$ & $15.62$ & $12.00$ \\
 $C_1^{\mathrm{av}}$ & $46.95$ & $11.09$ & $16.28$ & $9.74$ & $15.58$ \\
 $C_2^{\mathrm{av}}$ & $19.77$ & $6.16$ & $0.00$ & $8.63$ & $10.74$ \\
 $C_3^{\mathrm{av}}$ & $0.00$ & $0.45$ & $1.59$ & $0.00$ & $0.02$
\end{tabular}
\caption{List of critical velocities $\vc$[m/s], and fitting function coefficients $C_i$[$\mu$Pa\,s] for prismatic slip.
The critical velocities are given in units of m/s as well as in ratio to $\ct$.
The fits are only valid up to $0.99\ct$ and do not capture the asymptotic behavior for those metals/dislocations whose critical velocity is larger than this value.
Superscripts ``e'', ``s'', and ``av'' refer to ``edge'', ``screw'', and ``average'', respectively.
Furthermore, $\vc^{\text{av}}$ coincides with the smallest critical velocity for all dislocation characters $\vth$ within the prismatic slip system.}
\label{tab:coefficients_pris}
}
\end{table}


\begin{figure*}[h!t!b]
 \centering
 \includegraphics[width=0.368\textwidth,trim={1mm 2mm 2mm 2mm},clip]{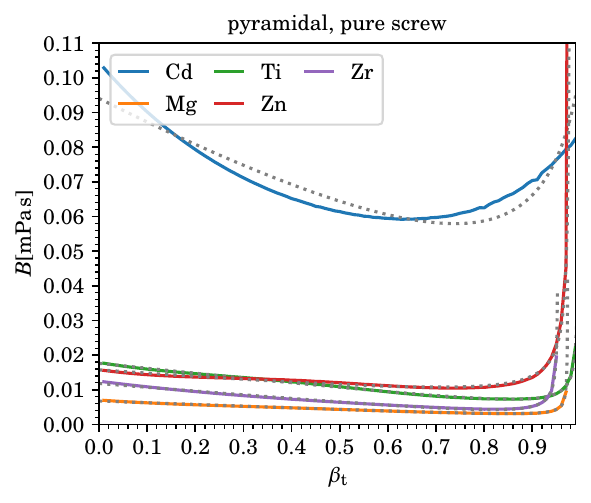}%
 \includegraphics[width=0.316\textwidth,trim={1.45cm 2mm 2mm 2mm},clip]{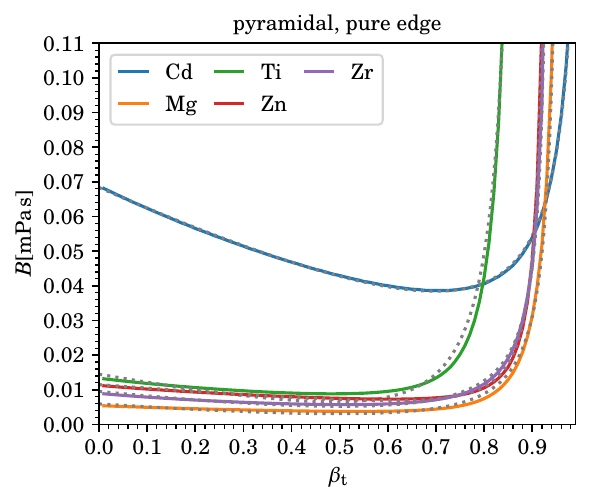}%
 \includegraphics[width=0.316\textwidth,trim={1.45cm 2mm 2mm 2mm},clip]{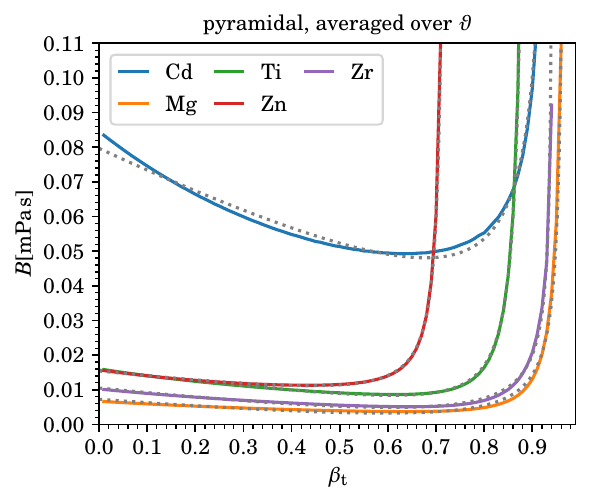}\vspace{-0.25cm}
\caption{%
We show the drag coefficient $B(\bt)$ from phonon wind for pure screw and edge dislocations as well as averaged over all character angles $\vth$ for pyramidal slip of 5 hcp metals.
Dashed lines represent the fitting functions and $\bt=v/\ct$.}
 \label{fig:drag-pyr}
\end{figure*}

\begin{table}[h!t!b]
{\renewcommand{\arraystretch}{1.2}
\small
\centering
 \begin{tabular}{c|c|c|c|c|c}
 pyramidal & Cd & Mg & Ti & Zn & Zr \\\hline\hline
 $\ct$ & 1486 & 3153 & 3118 & 2466 & 2350 \\
 $v_c^{\mathrm{e}}/\ct$ & 1.017 & 0.972 & 0.896 & 0.945 & 0.943 \\
 $v_c^{\mathrm{e}}$ & 1511 & 3065 & 2795 & 2332 & 2215 \\
 $C_0^{\mathrm{e}}$ & $68.37$ & $5.91$ & $14.48$ & $11.47$ & $9.51$ \\
 $C_1^{\mathrm{e}}$ & $62.25$ & $8.70$ & $21.63$ & $9.83$ & $12.96$ \\
 $C_2^{\mathrm{e}}$ & $31.95$ & $4.78$ & $0.00$ & $0.00$ & $12.51$ \\
 $C_3^{\mathrm{e}}$ & $0.50$ & $1.46$ & $5.48$ & $1.38$ & $0.51$
\\\hline\hline
 $v_c^{\mathrm{s}}/\ct$ & 1.061 & 0.974 & 0.996 & 0.980 & 0.953 \\
 $v_c^{\mathrm{s}}$ & 1577 & 3072 & 3105 & 2418 & 2239 \\
 $C_0^{\mathrm{s}}$ & $94.08$ & $6.75$ & $17.63$ & $15.74$ & $11.81$ \\
 $C_1^{\mathrm{s}}$ & $74.17$ & $4.95$ & $14.47$ & $9.63$ & $10.70$ \\
 $C_2^{\mathrm{s}}$ & $39.42$ & $0.62$ & $2.15$ & $4.23$ & $1.71$ \\
 $C_3^{\mathrm{s}}$ & $0.00$ & $0.00$ & $0.00$ & $0.04$ & $0.00$
\\\hline\hline
 $v_c^{\mathrm{av}}/\ct$ & 0.948 & 0.972 & 0.896 & 0.724 & 0.943 \\
 $v_c^{\mathrm{av}}$ & 1409 & 3065 & 2795 & 1786 & 2215 \\
 $C_0^{\mathrm{av}}$ & $79.66$ & $7.24$ & $15.67$ & $15.42$ & $10.51$ \\
 $C_1^{\mathrm{av}}$ & $60.93$ & $9.86$ & $14.32$ & $10.37$ & $13.09$ \\
 $C_2^{\mathrm{av}}$ & $24.11$ & $6.51$ & $1.45$ & $5.37$ & $9.07$ \\
 $C_3^{\mathrm{av}}$ & $0.83$ & $0.31$ & $1.25$ & $0.65$ & $0.02$
\end{tabular}
\caption{List of critical velocities $\vc$[m/s], and fitting function coefficients $C_i$[$\mu$Pa\,s] for pyramidal slip.
The critical velocities are given in units of m/s as well as in ratio to $\ct$.
The fits are only valid up to $0.99\ct$ and do not capture the asymptotic behavior for those metals/dislocations whose critical velocity is larger than this value.
Superscripts ``e'', ``s'', and ``av'' refer to ``edge'', ``screw'', and ``average'', respectively.
Furthermore, $\vc^{\text{av}}$ coincides with the smallest critical velocity for all dislocation characters $\vth$ within the pyramidal slip system.}
\label{tab:coefficients_pyr}
}
\end{table}


Based on the analysis of the previous section, the simplest form of a fitting function for the drag coefficient at fixed temperature, pressure, and dislocation character angle which captures its velocity dependence in the small $v$ as well as in the asymptotic regime $v\to\vc$ is given by
\begin{align}
 B(\vth)&\approx
 C_0(\vth) - C_1(\vth) x + C_2(\vth)\left(\frac1{(1-x^2)^{3/2}} - 1\right)
 \,, \nn\\
 x&=\frac{v}{v_c(\vth)} = \bt \frac{\ct}{v_c(\vth)}
 \,.\label{eq:fits_simple}
\end{align}
As is illustrated in Figure~\ref{fig:drag-Ni-theta} at the example of Ni at room temperature and ambient pressure for a number of dislocation character angles ranging from pure screw ($\vth=0$) to pure edge ($\vth=\pi/2$),
\eqnref{eq:fits_simple} is perfectly sufficient in some cases.
Corresponding fitting parameters (in units of $\mu$Pa$\,$s) and critical velocities --- all dependent on $\vth$ --- are listed in the figure legends and titles.


However, if $B$ shows a stronger $v$ dependence in the intermediate region, which is the case for a number of metals and slip systems, additional terms are required to improve the fits.
Candidates for such additional terms include of course any polynomial $x^k$ with $k\ge2$ or subleading divergences (which are always present), like $(1-x^2)^{-m}$ with $0<m<3/2$ or $\ln(1-x^2)$.
Our goal is to keep $B$ simple and the number of fitting parameters small.
Empirically, we find that adding only one additional term, $(1-x^2)^{-1/2}$, greatly improves the fits in most cases where \eqnref{eq:fits_simple} is insufficient.

Hence, better fits to the drag coefficient from phonon wind for a dislocation of fixed character angle $\vth$ are achieved using the function
\begin{align}
 B(\vth)&\approx
 C_0(\vth) - C_1(\vth) x + C_2(\vth)\left(\frac1{\sqrt{1-x^2}} - 1\right)
 \nn\\ &\quad 
 + C_3(\vth)\left(\frac1{(1-x^2)^{3/2}} - 1\right)
 \,, \nn\\
 x&=\frac{v}{v_c(\vth)} = \bt \frac{\ct}{v_c(\vth)}
 \,. \label{eq:fits}
\end{align}
Once again, it depends on the velocity in ratio to the critical velocity $\vc(\vth)$.
Note, that since $(nn)$ is a $3x3$ matrix, one always has three solutions for $\det(nn)=0$, and each can be represented as $\vc(\phi)$.
The branch with the smallest value for $\vc$ will lead to a divergence in $(nn)^{-1}$ first.
However, that solution need not always lead to a divergent drag coefficient since
subtle cancellations can lead to a finite dislocation field.
In this case, the second smallest value for $\vc$ is the true limiting velocity, see the recent review article \cite{Blaschke:2021vcrit} for details.

According values for $\vc$ corresponding to divergences in $B$, as well as the five fitting parameters $C_i$ for pure screw and edge dislocations, as well as for $B$ averaged over all dislocation character angles\footnote{
Averages were computed as mean values from $B(\vth)$ for 91 character angles ($0\le\vth\le\pi/2$ with $B(-\vth)=B(\vth)$) for fcc and hcp metals, and from 181 character angles ($-\pi/2<\vth\le\pi/2$) for bcc metals using~\cite{pydislocdyn}.
},
and each for the various metals computed (at room temperature and ambient pressure), are listed in Tables~\ref{tab:coefficients_cub}--\ref{tab:coefficients_pyr}.
Comparisons of these fits to the numerically computed results for $B$ are shown in Figures~\ref{fig:drag-cub}--\ref{fig:drag-pyr} as a function of velocity over effective transverse sound speed of the polycrystal, $\bt=v/\ct$.
Fits using \eqnref{eq:fits} can of course be derived for any other character angle $\vth$.
All numerical results presented here can be reproduced with the software of Ref.~\cite{pydislocdyn} developed by the present author.

With the exception of Ag, Au, and Cd, most metals shown here, have $B$ well below $0.04\,$mPas in the low velocity regime, 
for the most part due to lower values of their transverse sound speeds $\ct$ (see Tables~\ref{tab:coefficients_cub}, \ref{tab:coefficients_bas}).


Figures~\ref{fig:drag-Ni-theta}--\ref{fig:drag-pyr} show that $B(v,\vth$) at ambient temperature and pressure are well represented\footnote{
In our example of nickel, both equations, \eqref{eq:fits_simple}, \eqref{eq:fits}, yield exactly the same fit for pure edge dislocations.
In the case of screw dislocations, the 4-parameter equation \eqref{eq:fits} slightly improves an already decent fit by making use of the additional term $1/\sqrt{1-x^2}$, thereby changing also the values of the other 3 fitting parameters.
}
by \eqref{eq:fits} (or even \eqref{eq:fits_simple}), especially in light of the uncertainty in our current model for $B$ (which is hard to quantify):
For one, we considered only isotropic phonons, whose spectrum deviates from the true one especially in the high frequency regime.
Furthermore, we have neglected the dislocation core as well as the separation of dislocations into partials.
But also the uncertainties in the experimental (or computational) determination of the TOEC have a large effect on the accuracy of our present predictions.
We also need to stress that the present model is limited to the subsonic regime, i.e. $v<\vc(\vth)$ and $v<\ct$.
Furthermore, only straight dislocations moving at constant velocity were considered, i.e. the effect of acceleration or changes in shape are not (yet) considered.
Finally, the stress field required to reach velocities close to $\vc$ will likely lead to sizeable temperature and pressure gradients which would have to be considered in future improvements to $B$ as well.
A first attempt at incorporating the temperature dependence into $B$ is published in Ref. \cite{Blaschke:2021temperature}. 

Direct comparison of $B$ to experiments is limited to the low velocity regime (low meaning the viscous regime of $\bt\sim0.01$):
As was (in part) pointed out in Ref.~\cite{Blaschke:2018anis}, our predictions for $B(\bt\sim0.01)$ agree well with experimental results for Al (ranging from $\sim0.005\,$mPas to $\sim0.06\,$mPas, cf.~\cite{Hikata:1970,Gorman:1969,Parameswaran:1972}) and Cu (ranging from $\sim0.0079\,$mPas to $\sim0.08\,$mPas, cf.~\cite{Suzuki:1964,Zaretsky:2013,Stern:1962,Greenman:1967,Alers:1961}.
MD simulation results are in the range $\sim0.007$--$0.2\,$mPas for Al~\cite{Olmsted:2005,Yanilkin:2014,Cho:2017}, and $\sim0.016$--$0.022\,$mPas for Cu~\cite{Oren:2017,Wang:2008}.

Our predictions are lower than experimental results for Fe ($\sim0.34\,$mPas for edge and $\sim0.661\,$mPas for screw, cf.~\cite{Urabe:1975}) and Zn for both basal slip ($0.035\,$mPas for edge and $\sim0.034\,$mPas for screw, cf.~\cite{Pope:1969}) as well as for pyramidal slip ($0.27\,$mPas for edge and $\sim0.16\,$mPas for screw, cf.~\cite{Jassby:1977}).

Our drag coefficient for Mo is lower than the MD-simulation value of $\sim0.078\,$mPas for edge dislocations reported in~\cite{Weinberger:2010}.
Likewise, our drag coefficient for Ni, is lower than the MD-simulation results of $0.0321\,$mPas for edge dislocations reported in~\cite{Weinberger:2010}, and $\sim0.015\,$mPas for edge dislocations reported in~\cite{Bitzek:2004,Olmsted:2005,Daphalapurkar:2014}, albeit close to the latter.

\subsection*{Acknowledgements}

I thank D.~J. Luscher, C.~A. Bronkhorst, B. Feng, D.~L. Preston, and B.~A. Szajewski for related insightful discussions,
as well as B.~A. Szajewski for carefully reading the manuscript and the anonymous referees for their valuable comments.

This work was performed under the auspices of the U.S. Department of Energy under contract 89233218CNA000001.
In particular, the author is grateful for the support of the Advanced Simulation and Computing, Physics and Engineering Models Program.

\bibliographystyle{utphys-custom-twocol}
\bibliography{dislocations}

\end{document}